\documentstyle[12pt]{article}

\def\hybrid{\topmargin -20pt	\oddsidemargin 0pt
	\headheight 0pt	\headsep 0pt
	\textwidth 6.25in	
	\textheight 9.5in	
	\marginparwidth .875in
	\parskip 5pt plus 1pt	\jot = 1.5ex}

\hybrid

\def\baselinestretch{1.2}

\catcode`\@=11

\def\marginnote#1{}
\newcount\hour
\newcount\minute
\newtoks\amorpm
\hour=\time\divide\hour by60
\minute=\time{\multiply\hour by60 \global\advance\minute by-\hour}
\edef\standardtime{{\ifnum\hour<12 \global\amorpm={am}%
	\else\global\amorpm={pm}\advance\hour by-12 \fi
	\ifnum\hour=0 \hour=12 \fi
	\number\hour:\ifnum\minute<10 0\fi\number\minute\the\amorpm}}
\edef\militarytime{\number\hour:\ifnum\minute<10 0\fi\number\minute}

\def\draftlabel#1{{\@bsphack\if@filesw {\let\thepage\relax
   \xdef\@gtempa{\write\@auxout{\string
      \newlabel{#1}{{\@currentlabel}{\thepage}}}}}\@gtempa
   \if@nobreak \ifvmode\nobreak\fi\fi\fi\@esphack}
	\gdef\@eqnlabel{#1}}
\def\@eqnlabel{}
\def\@vacuum{}
\def\draftmarginnote#1{\marginpar{\raggedright\scriptsize\tt#1}}

\def\draft{\oddsidemargin -.5truein
	\def\@oddfoot{\sl preliminary draft \hfil
	\rm\thepage\hfil\sl\today\quad\militarytime}
	\let\@evenfoot\@oddfoot	\overfullrule 3pt
	\let\label=\draftlabel
	\let\marginnote=\draftmarginnote
   \def\@eqnnum{(\theequation)\rlap{\kern\marginparsep\tt\@eqnlabel}%
\global\let\@eqnlabel\@vacuum}  }

\def\preprint{\twocolumn\sloppy\flushbottom\parindent 2em
	\leftmargini 2em\leftmarginv .5em\leftmarginvi .5em
	\oddsidemargin -.5in	\evensidemargin -.5in
	\columnsep .4in	\footheight 0pt
	\textwidth 10.in	\topmargin  -.4in
	\headheight 12pt \topskip .4in
	\textheight 6.9in \footskip 0pt
	\def\@oddhead{\thepage\hfil\addtocounter{page}{1}\thepage}
	\let\@evenhead\@oddhead	\def\@oddfoot{}	\def\@evenfoot{} }

\def\numberbysection{\@addtoreset{equation}{section}
	\def\theequation{\thesection.\arabic{equation}}}

\def\underline#1{\relax\ifmmode\@@underline#1\else
	$\@@underline{\hbox{#1}}$\relax\fi}

\def\titlepage{\@restonecolfalse\if@twocolumn\@restonecoltrue\onecolumn
     \else \newpage \fi \thispagestyle{empty}\c@page\z@
	\def\thefootnote{\fnsymbol{footnote}} }

\def\endtitlepage{\if@restonecol\twocolumn \else \newpage \fi
	\def\thefootnote{\arabic{footnote}}
	\setcounter{footnote}{0}}  

\catcode`@=12
\relax

\def\figcap{\section*{Figure Captions\markboth
	{FIGURECAPTIONS}{FIGURECAPTIONS}}\list
	{Figure \arabic{enumi}:\hfill}{\settowidth\labelwidth{Figure
999:}
	\leftmargin\labelwidth
	\advance\leftmargin\labelsep\usecounter{enumi}}}
 \relax
\def\tablecap{\section*{Table Captions\markboth
	{TABLECAPTIONS}{TABLECAPTIONS}}\list
	{Table \arabic{enumi}:\hfill}{\settowidth\labelwidth{Table
999:}
	\leftmargin\labelwidth
	\advance\leftmargin\labelsep\usecounter{enumi}}}
 \relax
\def\reflist{\section*{References\markboth
	{REFLIST}{REFLIST}}\list
	{[\arabic{enumi}]\hfill}{\settowidth\labelwidth{[999]}
	\leftmargin\labelwidth
	\advance\leftmargin\labelsep\usecounter{enumi}}}
 \relax
\makeatletter
\newcounter{pubctr}
\def\publist{\@ifnextchar[{\@publist}{\@@publist}}
\def\@publist[#1]{\list
	{[\arabic{pubctr}]\hfill}{\settowidth\labelwidth{[999]}
	\leftmargin\labelwidth
	\advance\leftmargin\labelsep
	\@nmbrlisttrue\def\@listctr{pubctr}
	\setcounter{pubctr}{#1}\addtocounter{pubctr}{-1}}}
\def\@@publist{\list
	{[\arabic{pubctr}]\hfill}{\settowidth\labelwidth{[999]}
	\leftmargin\labelwidth
	\advance\leftmargin\labelsep
	\@nmbrlisttrue\def\@listctr{pubctr}}}
 \relax
\makeatother
\newskip\humongous \humongous=0pt plus 1000pt minus 1000pt

\newif\ifdtup

\relax

\def\s{\sigma}
\def\thefootnote{\fnsymbol{footnote}}
\def\be{\begin{equation}}
\def\ee{\end{equation}}
\def\ba{\begin{eqnarray}}
\def\ea{\end{eqnarray}}

\begin{document}
\renewcommand{\theequation}{\thesection.\arabic{equation}}
\newcommand{\beq}{\begin{equation}}
\newcommand{\eeq}[1]{\label{#1}\end{equation}}
\newcommand{\ber}{\begin{eqnarray}}
\newcommand{\eer}[1]{\label{#1}\end{eqnarray}}
\begin{titlepage}
\begin{center}

\hfill CERN--TH.7144/94\\
\hfill hep--th/9402016\\

\vskip .5in

{\large \bf O(2,2) TRANSFORMATIONS AND THE STRING \\
 GEROCH GROUP}

\vskip .8in

{\bf Ioannis Bakas}
\footnote{Permanent address: Department of Physics, University of
Crete,
GR--71409 Heraklion, Greece}
\footnote{e-mail address: BAKAS@SURYA11.CERN.CH}\\
\vskip .1in

{\em Theory Division\\
     CERN\\
     CH--1211 Geneva 23\\
     SWITZERLAND}\\

\vskip .4in

\end{center}

\vskip .6in

\begin{center} {\bf ABSTRACT } \end{center}
\begin{quotation}\noindent
The 1--loop string background equations with axion and dilaton fields
are shown to be integrable in four dimensions in the presence
of two commuting Killing symmetries and $\delta c = 0$.
Then, in analogy with reduced gravity, there is an infinite group
that acts on the space of solutions and generates
non--trivial string backgrounds from flat space. The usual $O(2,2)$
and $S$--duality transformations are just special cases of
the string Geroch group, which is infinitesimally identified
with the $O(2,2)$ current algebra. We also find an additional
$Z_{2}$ symmetry interchanging the field content of the dimensionally
reduced string equations. The method for constructing multi--soliton
solutions on a given string background is briefly
discussed.
\end{quotation}
\vskip1.0cm
CERN--TH.7144/94 \\
February 1994\\
\end{titlepage}
\vfill
\eject

\def\baselinestretch{1.2}
\baselineskip 16 pt
\setcounter{section}{0}
\section{\bf Introduction}
\noindent
In realistic string models the target space is described by $M_{4}
\times
K$, where $M_{4}$ is a 4--dim space with signature $- + + +$ and $K$
is
some internal space, which is usually represented by a conformal
field
theory, so that the total central charge
of the string theory is at its critical value. The
4--dim Minkowski space provides a particular choice for $M_{4}$, but
it is
well known by now that there are many other choices
with non--trivial dilaton and antisymmetric tensor (or axion)
fields that are compatible with local scale invariance
quantum mechanically [1]. A variety of cosmological and black hole
type
solutions have been constructed explicitly using conformal
field theory techniques, duality symmetries, and $O(d, d)$
transformations in general; the list of references [2--10] is
indicative, but by no means complete.
Although the solution generating techniques we have available at the
moment are quite interesting for further study, they appear to be
rather restrictive in many ways. It would be interesting to find new
(and perhaps complementary) methods for constructing general classes
of
string vacua and study the transition between them in a systematic
way.

The string background equations with axion and dilaton fields
provide a natural generalization of the vacuum Einstein equations
in $M_{4}$. Finding their most general solution is a hopeless
problem. The best we can hope for our solution generating techniques
will be the construction of non--trivial string backgrounds with a
certain degree of symmetry, which effectively reduces the
dimensionality of the corresponding string equations. In the
mini--superspace approach the string background equations are reduced
to a 1--dim non--linear system, which is of interest for some
cosmological applications, but quite restrictive otherwise. In this
paper we adopt a midi--superspace approach, which preserves the field
theory aspects of the problem, while making the construction of a
large class of solutions still possible.

We consider 4--dim backgrounds with two commuting Killing symmetries
and show that the dimensionally reduced 1--loop string equations
are in fact integrable. We also find an infinite dimensional
symmetry on the space of solutions, which is infinitesimally
described
by the $SL(2,R) \times SL(2,R) \simeq O(2,2)$ current algebra.
Then, finite group elements (i.e. elements further away from the
vicinity of the identity) can be used to generate
(at least formally) non--trivial
string backgrounds from flat space or from other known solutions
with two Killing symmetries. We call this new infinite dimensional
symmetry the string Geroch group, because for constant axion and
dilaton fields it becomes the infinite group of reduced vacuum
Einstein gravity generated by the $SL(2,R)$ current algebra [11--16].
In the string case, the usual $O(2,2)$ and $S$--duality
transformations
can be explicitly identified with certain modes of the $O(2,2)$
Geroch group. As we will see later, we have to go beyond the zero
modes to achieve this in the axion formalism.

It should be noted that the explicit construction
of new solutions with the aid of the Geroch group is not an easy
project to complete,
because in the general case we have to solve the associated
Riemann-Hilbert problem, which is described by a singular integral
equation. The class of solutions which are more accessible to
explicit calculations, however,
is that of the (multi)--soliton solutions on any
given string background. It should be emphasized that soliton
solutions correspond to special elements of the full
string Geroch group
and cannot be obtained by the usual $O(2,2)$ or
$S$--duality transformations. Many
non--trivial solutions we already know in reduced vacuum Einstein
gravity admit a solitonic interpretation on the background of much
simpler solutions; for example, the exterior of a rotating black hole
can be described as a double stationary soliton on flat space [14].
Hence, it is natural to expect that straightforward generalization of
the
Belinski-Sakharov method will produce an infinite (but discrete) set
of new 4--dim string solutions descending from known ones.
Multi--soliton excitations in dimensionally reduced string theory
will be characterized by two positive integers $(n, m)$, referring to
the soliton numbers associated with the two $SL(2, R)/U(1)$
$\s$--models that appear in the effective 2--dim description of the
string background equations.

The dimensional reduction will be perfomed in the case that both
Killing symmetries of $M_{4}$ are space--like. This is merely done
for practical reasons, fixing the notation and sign conventions
throughout the calculations. We note, however, that the whole
discussion can be easily generalized to the case of one space--like
and one time--like Killing symmetry, as well as to 4--dim spaces with
Euclidean or $(2,2)$ signature. A few remarks will be made about
these
other cases, but the generalization will not be discussed in detail.

The remaining part of this paper is organized as follows. In section
2
we consider the string background equations with zero cosmological
constant (i.e. $\delta c = 0$)
and perform the dimensional reduction in the presence of two
commuting Killing symmetries. In section 3 we describe the
infinitesimal action of the string Geroch group and identify its
generators with the $O(2,2)$ current algebra. The embedding of the
$O(2,2)$ and $S$--duality transformations will be discussed in
detail.
In section 4 we find that every solution with two space--like
Killing symmetries
possess a ``mirror image" that is obtained by interchanging the field
content of the two $SL(2, R)/U(1)$ $\s$--models
that appear in the formalism. Any two solutions of
this kind admit different space--time interpretations, in general.
In section 5 we review the general method for constructing
(multi)--soliton solutions and describe their form around
simple string background. Meron--like solutions are also briefly
discussed. Finally, in section 6 we present the
conclusions and directions for future work.

\section{\bf Reduced String Background Equations}
\noindent
String propagation in a non--trivial background is described by a
generalized 2--dim non--linear $\s$--model, which in the conformal
gauge, ${e}^{2 \s} dz d \bar{z}$, assumes the form
\be
S = {1 \over 4 \pi {\alpha}^{\prime}} \int dz d \bar{z}
\left((G_{\mu \nu}^{(\s)} (X) + B_{\mu \nu} (X))
\partial X^{\mu} \bar{\partial} X^{\nu} - 4 {\alpha}^{\prime}
(\partial \bar{\partial} \s) \Phi \right).
\ee
Here $G_{\mu \nu} (X)$,
$B_{\mu \nu} (X)$ and $\Phi (X)$ are the target space metric,
the antisymmetric tensor and the dilaton fields, respectively.
It is well known
that these fields have to satisfy certain consistency conditions so
that the quantum theory possesses local scale invariance. In
particular, at the 1--loop level in the coupling constant
${\alpha}^{\prime}$ (inverse string tension), the vanishing
conditions for the beta functions ${\beta}^{\Phi}$,
${\beta}_{\mu \nu}^{G}$ and ${\beta}_{\mu \nu}^{B}$ are [1]
\be
4 {(\nabla \Phi)}^{2} - 4 {\nabla}^{2} \Phi - R^{(4)}[G^{(\s)}] +
{1 \over 12}
H^{2} = 0,
\ee
\be
R_{\mu \nu}^{(4)} [G^{(\s)}] - {1 \over 4} {H_{\mu}}^{\lambda \s}
H_{\nu \lambda \s}
+ 2 {\nabla}_{\mu} {\nabla}_{\nu} \Phi = 0,
\ee
\be
{\nabla}_{\lambda} {H_{\mu \nu}}^{\lambda} - 2 ({\nabla}_{\lambda}
\Phi)
{H_{\mu \nu}}^{\lambda} = 0,
\ee
respectively, where $H_{\mu \nu \lambda} = 3 {\nabla}_{[\mu}
B_{\nu \lambda ]}$ is the field strength of the antisymmetric tensor
field. The right--hand side of eq. (2.2) is set equal to zero
because we assume that the central charge deficit
$\delta c$ (cosmological constant) is zero to first order in
${\alpha}^{\prime}$. In other words we make the assumption that
string propagation takes place in $M_{4} \times K$
with $c(M_{4}) = 4 + \delta c$ and $c(K) = 22 - \delta c$, where
$\delta c = {\cal O}({({\alpha}^{\prime})}^{2})$.
The details of the internal space $K$
will not be needed for the purposes of the present work.
We also recall for completeness that eq. (2.2) is a consequence of
eqs. (2.3) and (2.4), using the Bianchi identities.

It is convenient to rewrite the 4--dim string background equations
in a form that will later lead to a separation of variables upon
dimensional reduction. After a conformal rescaling of the metric,
\be
G_{\mu \nu} = e^{- 2 \Phi} G_{\mu \nu}^{(\s)},
\ee
we pass from the $\s$--model frame to the Einstein frame, in which
the
string equations (2.2)--(2.4) are equivalently described by the
classical equations of motion of the effective action
\be
S_{eff} = \int_{M_{4}} d^{4} X \sqrt{-\det G} \left( R^{(4)}[G] -
2 {(\nabla \Phi)}^{2}
- {1 \over 12} e^{-4 \Phi} H^{2} \right).
\ee
In four dimensions we can also trade $B_{\mu \nu} (X)$ for a scalar
field
$b(X)$ by duality. The axion field can be consistently defined
in the Einstein frame as follows,
\be
{\partial}_{\mu} b = {e^{-4 \Phi} \over 6} \sqrt{-\det G} ~
{{\epsilon}_{\mu}}^{\nu \rho \s} H_{\nu \rho \s},
\ee
where ${\epsilon}_{0123} = 1$.
In $M_{4}$ with signature $-+++$ we may further define the complex
conjugate fields
\be
S_{\pm} (X) = b \pm i e^{-2 \Phi}.
\ee
The axion formalism should be introduced directly to the classical
equations of motion and not to the effective action (2.6), since
otherwise sign discrepancies will arise. Then, in the Einstein frame,
the 1--loop effective action assumes the form
\be
S_{eff} = \int_{M_{4}} d^{4} X \sqrt{-\det G} \left( R^{(4)}[G] +
2 G^{\mu \nu} {{\partial}_{\mu} S_{+} {\partial}_{\nu} S_{-}
\over {(S_{+} - S_{-})}^{2}} \right),
\ee
which describes a 4--dim $SL(2,R)/U(1)$ non--linear $\s$--model
coupled
to gravity. In spaces with Minkowski signature,
like $M_{4}$, this $\s$--model is
Euclidean since its target space has signature $++$. On the contrary,
in 4-dim spaces with Euclidean or $(2,2)$ signature the $\s$--model
is
Lorentzian and the relevant $\s$--model variables are
$S_{\pm} = b \pm e^{-2 \Phi}$.

It will be quite useful for our purposes to adopt an alternative
description of the $SL(2,R)/U(1)$ $\s$--model by introducing the
symmetric $2 \times 2$ matrix
\be
\lambda (X) = e^{2 \Phi} \left( \begin{array}{ccc}
1 &  & b\\
  &  &  \\
b &  & b^{2} + e^{-4 \Phi}
                \end{array} \right)
\ee
with $\det \lambda = 1$, so that the 4--dim effective string
action is equivalently written as
\be
S_{eff} = \int_{M_{4}} d^{4} X \sqrt{-\det G} \left(R^{(4)}[G] -
{1 \over 4} \mbox{Tr} (J_{\mu} J^{\mu}) \right),
\ee
where
\be
J_{\mu} (X) = {\lambda}^{-1} {\partial}_{\mu} \lambda.
\ee
We mention for completeness that for the Lorentzian $SL(2,R)/U(1)$
$\s$--model the matrix element $b^{2} + e^{-4 \Phi}$ has to be
replaced by $b^{2} - e^{-4 \Phi}$,
$b$ also being the axion field, so that
$\det \lambda = -1$ in that case.

The string background equations, which follow from the action (2.12),
will be dimensionally reduced in the presence of two commuting
space--like Killing symmetries. We assume, therefore, that
the components of
$G_{\mu \nu}$ depend only on $X^{0}$ and $X^{1}$. We also make the
extra assumption of orthogonal transitivity, which
physically means that $M_{4}$ possess a reflection symmetry under
$(X^{2}, X^{3}) \rightarrow (-X^{2}, -X^{3})$. This implies that
the metric has the block diagonal form
\be
{ds}^{2} = h_{ij} (X^{0}, X^{1}) dX^{i} dX^{j} + g_{AB} (X^{0},
X^{1})
dX^{A} dX^{B},
\ee
with $G_{iA} = 0$. Here $i, j$ take the values 0 or 1, while $A, B$
take 2 or 3. Also, the 2--dim metric $g$ has Euclidean signature.
Since the $h$--part of the metric can always be brought into a
conformally flat form, we may rewrite the line element (2.13) as
\be
{ds}^{2} = - f(\eta, \xi) d \eta d \xi + g_{AB} (\eta, \xi)
dX^{A} dX^{B},
\ee
where $f(\eta, \xi)$ is the conformal factor and $\eta, \xi$ are the
light--cone coordinates
\be
\eta = {1 \over 2} (X^{0} - X^{1}), ~~~~
\xi = {1 \over 2} ( X^{0} + X^{1}).
\ee

As for the axion and dilaton fields, we also assume that
they depend only on
$\eta$ and $\xi$. Using eq. (2.7) we immediately see that
\be
B_{23} = -B_{32} \equiv B(\eta, \xi),
\ee
while all other components of
the antisymmetric tensor field vanish
in this case. Then, in the light--cone
variables, the relations (2.7) can be written as
\be
{\partial}_{\xi} b = {e^{-4 \Phi} \over \sqrt{\det g}} ~
{\partial}_{\xi} B, ~~~ {\partial}_{\eta} b =
- {e^{-4 \Phi} \over \sqrt{\det g}} ~ {\partial}_{\eta} B.
\ee
It will turn out that the
sector of the string background equations described by this ansatz
is an integrable 2--dim system, generalizing a similar situation
encountered in vacuum Einstein gravity [12--16].

The string background equations can be easily cast into the form
\be
R_{\mu \nu}^{(4)}[G] = {1 \over 4} \mbox{Tr} (J_{\mu} J_{\nu}),
\ee
\be
{\nabla}_{\mu} J^{\mu} = 0,
\ee
after eliminating the Ricci scalar curvature term by contraction. For
the special class of metrics (2.12) we are considering here, the
dimensional reduction of the string equations yields
\be
R_{iA}^{(4)}[G] = 0,
\ee
\be
R_{AB}^{(4)}[G] = 0,
\ee
\be
R_{ij}^{(4)}[G] = {1 \over 4} \mbox{Tr} (J_{i} J_{j}),
\ee
\be
{\nabla}_{i} J^{i} + {{\partial}_{i} \sqrt{\det g} \over
\sqrt{\det g}} J^{i} = 0.
\ee
It can be verified that eq. (2.20) is just an identity and has no
field content. All the information about the dynamics of the field
variables is contained in eqs. (2.21)--(2.23), which we now analyse
case by case in the light--cone variables.

Equation (2.21) does not depend on the axion--dilaton
system and hence
it is identical to the same equation of reduced vacuum Einstein
gravity. Explicit calculation shows that it is equivalent to the
following continuity equation for the $2 \times 2$ metric $g$,
\be
{\partial}_{\eta} (\sqrt{\det g} ~ g^{-1} {\partial}_{\xi} g) +
{\partial}_{\xi} (\sqrt{\det g} ~ g^{-1} {\partial}_{\eta} g) = 0
\ee
and
\be
{\partial}_{\eta} {\partial}_{\xi} (\sqrt{\det g}) = 0.
\ee
Equation (2.23) yields a similar equation for
the $2 \times 2$ symmetric
matrix $\lambda$, namely
\be
{\partial}_{\eta} (\sqrt{\det g} ~ {\lambda}^{-1} {\partial}_{\xi}
\lambda) + {\partial}_{\xi} (\sqrt{\det g} ~ {\lambda}^{-1}
{\partial}_{\eta} \lambda) = 0.
\ee

At this point there appears to be a slight difference between eqs.
(2.24)
and (2.26) in that $\det g \neq \det \lambda = 1$. We note, however,
that
if we define ${\lambda}^{\prime} = \sqrt{\det g} ~ \lambda$ so that
$\det {\lambda}^{\prime} = \det g$, the continuity equation for
${\lambda}^{\prime}$ is identical to that of $\lambda$ thanks to the
Laplacian condition (2.25). As a result, the field variables
$g$, $\lambda$ satisfy the classical equations of motion of two
$SL(2,R)/U(1)$ non--linear $\s$--models in two dimensions,
modified by the presence of the $\sqrt{\det g}$ factor in their
currents. Modified $\s$--models of this type are usually called
Ernst models. They first arose in Geroch's treatment of reduced
gravity [12--14]
and were subsequently studied in detail by Ernst and collaborators,
in the vacuum and electrovacuum cases [15].
In $M_{4}$ with two space--like Killing symmetries, both $\s$--models
have Euclidean signature. In section 4 we will use this fact to
produce new solutions by interchanging their field content.

The 2--dim wave equation (2.25) implies that $\sqrt{\det g}$ can be
written as a sum of two arbitrary functions, one depending on
$\eta$ and the other on $\xi$ only. Let $\alpha (\eta, \xi)$ and
$\beta (\eta, \xi)$ be a pair of conjugate solutions such that
\be
{\partial}_{\xi} \alpha = {\partial}_{\xi} \beta, ~~~~
{\partial}_{\eta} \alpha = - {\partial}_{\eta} \beta.
\ee
Then, without loss of generality, we may choose
\be
\alpha = \xi + \eta, ~~~~ \beta = \xi - \eta,
\ee
since the form of the line element (2.14) remains invariant under the
transformation. The differential equations we are going to derive
for the conformal factor $f(\eta, \xi)$ simplify in the frame
\be
\sqrt{\det g} = \alpha = \xi + \eta.
\ee
Its essential properties remain the same, however, while the more
general picture can be easily described by a simple transformation.

The conditions on the conformal factor $f$ follow from eq. (2.22),
which describes its dependence on $g$ and $\lambda$. We first find
\be
R_{ij}^{(4)}[G] = R_{ij}^{(2)}[h] - {\nabla}_{i} {\nabla}_{j}
(\log \sqrt{\det g}) - {1 \over 4}
\mbox{Tr} \left( (g^{-1} {\partial}_{i}
g)(g^{-1} {\partial}_{j} g) \right).
\ee
Then, in the light--cone metric (2.14), since
\be
R_{\eta \eta}^{(2)} = 0 = R_{\xi \xi}^{(2)},
\ee
we obtain the following first--order conditions
\be
{\partial}_{\xi} (\log f) = - {1 \over \alpha} + {\alpha \over 4} ~
\mbox{Tr} \left( {(g^{-1} {\partial}_{\xi} g)}^{2} +
{({\lambda}^{-1} {\partial}_{\xi} \lambda)}^{2} \right),
\ee
\be
{\partial}_{\eta} (\log f) = - {1 \over \alpha} + {\alpha \over 4} ~
\mbox{Tr} \left( {(g^{-1} {\partial}_{\eta} g)}^{2} +
{({\lambda}^{-1} {\partial}_{\eta} \lambda)}^{2} \right),
\ee
provided that eq. (2.29) is satisfied. In a more general frame, the
first term on the right--hand side has to be replaced by
$\partial[\log (\partial(\log \sqrt{\det g}))]$ and the overall
coefficient of the trace term by
$1/ 4 \partial(\log \sqrt{\det g})$; the
derivatives are taken with respect to $\xi$ or $\eta$ when referring
to the generalization of eq. (2.32) or (2.33), respectively.

This simple system of equations for $\log f$ can be
easily integrated, once a pair of solutions $(g, \lambda)$
for the two $SL(2,R)/U(1)$ Ernst $\s$--models
is known. This system is also
compatible with the other equations following from the
dimensional reduction. It can be verified that their compatibility
is fully encoded in the $(\eta \xi)$--component of eq. (2.22), which
is the last one to examine. For this we also have to make use of the
expression
\be
R_{\eta \xi}^{(2)} = -  {\partial}_{\eta} {\partial}_{\xi}
(\log f)
\ee
for the 2--dim curvature.
Hence, we find that in this case the problem of generating new
solutions
to the string background equations reduces to the solution of two
$SL(2,R)/U(1)$ Ernst $\s$--models.
The two $\s$--models are essentially decoupled because
$\sqrt{\det g}$ satisfies the wave equation (2.25). This
decoupling is manifest in the special coordinate system (2.28),
which eliminates the determinant degree of freedom.
Our formalism also reproduces the classical equations
of reduced gravity for constant matrix $\lambda$.

One might think that it would be possible to choose the condition
$\sqrt{\det g} = 1$ and still be able to describe a large class of
solutions to the string background equations. There is a
uniqueness theorem, however, which was proved for reduced
gravity a long time ago [11] and can be easily extended to the string
case. In particular, for $\sqrt{\det g} = 1$, the only physical
solution is the flat Minkowski space with trivial dilaton and
axion fields. The point is that
although there are infinitely many solutions
of the ordinary $SL(2,R)/U(1)$ $\s$--model (to which the Ernst
equation reduces for $\sqrt{\det g} = 1$), the additional conditions
on the conformal factor $f(\eta, \xi)$ reduce the number of physical
possibilities to the trivial one. This will be used later to
clarify the meaning of the string Geroch group.

\section{\bf On the String Geroch Group}
\setcounter{equation}{0}
\noindent
In $M_{4}$ with orthogonal transitivity and two commuting
space--like Killing symmetries, the string background equations
are described by the 2--dim system of Ernst equations (2.24)--(2.26),
plus a simple system of first--order differential equations for the
conformal factor. There are six degrees of freedom in the problem,
but because of the decoupling that
essentially occurs in their dynamics the
target space integrability of string theory boils down to the
integrability of the $SL(2,R)/U(1)$ Ernst $\s$--model. This is a
well studied model in reduced gravity and exhibits infinitely many
symmetries, in analogy with ordinary 2--dim non--linear $\s$--models.
We will briefly review some of its integrability aspects and then
identify the known $O(2,2)$ and $S$--duality transformations of
string theory in the context of the corresponding string Geroch
group, which is much larger.

Consider the Ernst $\s$--model
\be
{\partial}_{\xi}(\alpha g^{-1} {\partial}_{\eta} g) +
{\partial}_{\eta}(\alpha g^{-1} {\partial}_{\xi} g) = 0,
\ee
where $\alpha = \sqrt{\det g}$, but without necessarily making the
same choice of coordinates as in (2.28).
It is convenient for our purposes to
adopt a complexified description of the model and impose the
reality conditions at the end. For this we first introduce a twist
potential matrix $\psi(\eta, \xi)$ such that
\be
{\partial}_{\xi} \psi = \alpha \epsilon g^{-1} {\partial}_{\xi} g,
{}~~~ {\partial}_{\eta} \psi = - \alpha \epsilon g^{-1}
{\partial}_{\eta} g,
\ee
where $\epsilon$ is the usual antisymmetric matrix
\be
\epsilon = \left( \begin{array}{ccc}
0  &   & 1\\
   &   &  \\
-1 &   & 0        \end{array} \right).
\ee
The existence of $\psi$ is guaranteed by the integrability condition
of the system (3.2), which is nothing else but the $\s$--model
equation
(3.1). We then introduce the so--called Ernst potential matrix [15]
\be
E = g + i \psi
\ee
and reformulate the problem in terms of $E$.

It can be verified that the linearization system of the theory (3.1)
is given by
\be
{\partial}_{\xi} F(l) = {l \over 1 - 2l(\beta + \alpha)} ~
({\partial}_{\xi} E) i \epsilon F(l),
\ee
\be
{\partial}_{\eta} F(l) = {l \over 1 - 2l(\beta - \alpha)} ~
({\partial}_{\eta} E) i \epsilon F(l),
\ee
where $F(l) = F(\eta, \xi ; l)$ is a $2 \times 2$ matrix depending
on a spectral parameter $l$ with $F(l=0) = 1$. Here $\alpha$, $\beta$
is the pair of conjugate solutions (2.27) of the 2--dim wave
equation for $\sqrt{\det g}$. This linearization is sufficient to
establish the integrability of the $SL(2,R)/U(1)$ Ernst $\s$--model.
We also note some properties of $g$ and $\psi$, which are useful for
the calculations,
\be
g \epsilon g = {\alpha}^{2} \epsilon, ~~~~ \psi - {\psi}^{t} =
2 \beta \epsilon,
\ee
where ${\psi}^{t}$ denotes the transpose of $\psi$.

The model exhibits a hidden symmetry of (non--local) transformations,
in close analogy with 2--dim principal chiral models [17].
Following [16], we define
\be
{\delta}_{T}E = - {1 \over l} ~ \left(F(l)T{F(l)}^{-1} - T\right)
(i \epsilon),
\ee
where $l$ is taken to be real and $T$ is an element of the
$SL(2,R)$ algebra,
\be
T = {\epsilon}_{+} T_{-} + {\epsilon}_{-} T_{+} + {\epsilon}_{0}
T_{0},
\ee
with arbitrary infinitesimal real parameters ${\epsilon}_{\pm}$,
${\epsilon}_{0}$. It follows that
\be
{\delta}_{T}g = - {1 \over l} ~ \mbox{Re}
\left( F(l)T{F(l)}^{-1} i \epsilon
\right)
\ee
is a symmetric $2 \times 2$ matrix, which defines the infinitesimal
action of the Geroch group on the space of solutions of the model.
An important property of this transformation is
\be
{\delta}_{T}\alpha = 0 = {\delta}_{T}\beta
\ee
and hence $\sqrt{\det g}$ remains invariant.

Introducing the loop expansion of the variation
\be
{\delta}_{T} = \sum_{n=0}^{\infty} l^{n} {\delta}_{T}^{(n)} ~ ,
\ee
it has been proven in the literature [15--17] (see also [12, 13] for
alternative derivations) that
\be
[{\delta}_{T}^{(n)} ~ , ~ {\delta}_{T^{\prime}}^{(m)}] =
{\delta}_{[T, T^{\prime}]}^{(n+m)} ~ ,
\ee
which is an $SL(2,R)$ current algebra. The key formula in deriving
this
result is
\be
{\delta}_{T}F(t) = {t \over t-l} ~ \left( F(l)T{F(l)}^{-1} -
F(t)T{F(t)}^{-1} \right) F(t).
\ee
So far we have actually described only half of the modes since both
$m$, $n$ are $\geq 0$. The negative modes can be appended as well
by a slight modification of the formalism, but it is well known that
the latter give rise to trivial (gauge) transformations on $g$.

In the string case, where we have two such Ernst $\s$--models,
the corresponding string Geroch group is generated by the
$SL(2,R) \times SL(2,R) \simeq O(2,2)$ current algebra. However,
only its non--negative modes will lead to non--trivial
transformations
of $g$ and $\lambda$, leaving the string equations invariant.
The variation of the conformal factor $f$ follows immediately from
the transformation of $g$ and/or $\lambda$, so that it still
satisfies
eqs. (2.32), (2.33) (or their generalization discussed earlier).

In the following we need some more explicit expressions for the
variation of $g$ (and $\lambda$). One may use for this the linearized
system of equations (3.5), (3.6) writing $F(l)$ as a
path--ordered exponential of gauge connections depending on $E$.
The results will be valid off shell as well if we use only one
of the two equations, say eq. (3.5), in analogy with a similar
derivation in 2--dim principal chiral models [17].
With this in mind,
expanding the path--ordered exponential form of $F(l)$ in powers
of $l$ around $l=0$, we obtain the following result
\be
{\delta}_{T}^{(0)}g = [g \epsilon ~ , ~ T] \epsilon
\ee
to zeroth order. To first order we have
\be
{\delta}_{T}^{(1)}g = [g \epsilon ~ , ~ T \psi \epsilon -
2 \psi \epsilon T + {\psi}^{t} \epsilon T] \epsilon,
\ee
using the properties (3.7), and so on. Unlike the variation
${\delta}_{T}^{(0)}g$, which is local, ${\delta}_{T}^{(1)}g$ is
non--local in $g$ since the right--hand side involves the
twist potential matrix $\psi$. This will be very important later on.
These transformations apply equally well to the $g$ and $\lambda$
sectors of the dimensionally reduced string background equations.
We note for completeness that since $g$ is a symmetric matrix,
$g \epsilon$ is traceless; in these variables the variations
(3.15) and (3.16) look closer to the similar variations of
ordinary $\s$--models.

To justify our claim that the usual $O(2,2)$ and $S$--duality
transformations are just special cases of the string Geroch group,
we have to formulate them in the Einstein frame (2.5), on which
our previous discussion was based.

Recall that any element of the $O(2,2)$ group can be represented
in the form
\be
D = \left( \begin{array}{ccc}
D_{1}  &    & D_{2}\\
       &    &      \\
D_{3}  &    & D_{4} \end{array} \right), ~~~~
D \left( \begin{array}{ccc}
0  &   & 1\\
   &   &  \\
1  &   & 0 \end{array} \right) D^{t} =
\left( \begin{array}{ccc}
0  &   & 1\\
   &   &  \\
1  &   & 0 \end{array} \right),
\ee
where $D_{1}, \cdots , D_{4}$ are $2 \times 2$ matrices otherwise
arbitrary. Expanding $D$ to first order in an infinitesimal
parameter $\epsilon$, we obtain
\be
D_{1} = 1 + \epsilon d_{1}, ~~ D_{2} = \epsilon d_{2}, ~~
D_{3} = \epsilon d_{3}, ~~ D_{4} = 1 + \epsilon d_{4},
\ee
where the Lie algebra elements $d_{1}, \cdots , d_{4}$ satisfy the
$O(2,2)$ constraints
\be
d_{2} = - {d_{2}}^{t}, ~~~ d_{3} = - {d_{3}}^{t} , ~~~
d_{4} = -{d_{1}}^{t}.
\ee
The decomposition of $O(2,2)$ into two commuting $SL(2,R)$ algebras
can
be described in this representation by choosing the basis elements
\be
d_{+} = \left( \begin{array}{cccc}
0  & 1 & 0 & 0\\
0  & 0 & 0 & 0\\
0  & 0 & 0 & 0\\
0  & 0 &-1 & 0  \end{array}  \right), ~~
d_{-} = \left( \begin{array}{cccc}
0 & 0 & 0 & 0\\
1 & 0 & 0 & 0\\
0 & 0 & 0 &-1\\
0 & 0 & 0 & 0  \end{array}  \right), ~~
d_{0} = \left( \begin{array}{cccc}
1 & 0 & 0 & 0\\
0 &-1 & 0 & 0\\
0 & 0 &-1 & 0\\
0 & 0 & 0 & 1   \end{array}  \right)
\ee
for one of them and
\be
d_{+} = \left( \begin{array}{cccc}
0 & 0  &  0 &-1\\
0 & 0  &  1 & 0\\
0 & 0  &  0 & 0\\
0 & 0  & 0  & 0  \end{array} \right), ~~
d_{-} = \left( \begin{array}{cccc}
0 & 0 & 0 & 0\\
0 & 0 & 0 & 0\\
0 & 1 & 0 & 0\\
-1& 0 & 0 & 0   \end{array}  \right), ~~
d_{0} = \left( \begin{array}{cccc}
1  & 0 & 0 & 0\\
0  & 1 & 0 & 0\\
0  & 0 &-1 & 0\\
0  & 0 & 0 &-1  \end{array}  \right)
\ee
for the other $SL(2,R)$ factor, so that $[d_{+} , d_{-}] = d_{0}$
and $[d_{0} , d_{\pm}] = \pm 2 d_{\pm}$.

It is known that 4--dim string backgrounds with two
commuting Killing fields exhibit an $O(2,2)$ symmetry, which
leaves the 1--loop beta function equations invariant [8, 9]. For the
special class of block diagonal metrics (2.14), we consider the
$\s$--model variables
\be
{\cal G}_{AB}^{(\s)}(\eta, \xi) = g_{AB}^{(\s)} + {\epsilon}_{AB} B,
{}~~~ f^{(\s)}(\eta, \xi) = e^{2 \Phi} f,
\ee
with $B$ defined as in eq. (2.16) and ${\epsilon}_{23} = 1$.
Then, the invariant action of an
arbitrary $O(2,2)$ group element $D$, (3.17), is described by the
transformation
\be
{\cal G}^{(\s)} \rightarrow {\tilde{{\cal G}}}^{(\s)} =
(D_{1} {\cal G}^{(\s)} + D_{2})
{(D_{3} {\cal G}^{(\s)} + D_{4})}^{-1},
\ee
\be
f^{(\s)} \rightarrow {\tilde{f}}^{(\s)} = f^{(\s)},
\ee
\be
e^{2 \Phi} \rightarrow e^{2 \tilde{\Phi}} = e^{2 \Phi}
\sqrt{{\det {\tilde{g}}^{(\s)} \over \det g^{(\s)}}}.
\ee
Note that eq. (3.25) implies that in the Einstein frame
\be
\sqrt{\det \tilde{g}} = \sqrt{\det g},
\ee
which is a common property with the Geroch group (cf. eq. (3.11)).
This motivates us to examine whether the remaining $O(2,2)$
transformations are also part of the string Geroch group.

The infinitesimal form of the action (3.23) is
\be
\delta {\cal G}^{(\s)} = \epsilon (d_{1} {\cal G}^{(\s)} +
{\cal G}^{(\s)} {d_{1}}^{t} + d_{2} -
{\cal G}^{(\s)} d_{3} {\cal G}^{(\s)}),
\ee
where $\epsilon$ is a collective infinitesimal parameter
(not to be confused with the antisymmetric matrix
(3.3)). Taking the
symmetric and antisymmetric parts of this variation we obtain in
the $\s$--model frame the expressions for $\delta g_{AB}^{(\s)}$
and $\delta B$ respectively. The results of the calculation will be
described separately for the two $SL(2,R)$ subalgebras to which
$O(2,2)$ is decomposed.

(i). For the $SL(2,R)$ subalgebra (3.20) we obtain
\be
\delta g_{22}^{(\s)} = 2 ({\epsilon}_{-} g_{23}^{(\s)} +
{\epsilon}_{0} g_{22}^{(\s)}),
\ee
\be
\delta g_{33}^{(\s)} = 2({\epsilon}_{+} g_{23}^{(\s)} -
{\epsilon}_{0} g_{33}^{(\s)}),
\ee
\be
\delta g_{23}^{(\s)} = {\epsilon}_{+} g_{22}^{(\s)} +
{\epsilon}_{-} g_{33}^{(\s)},
\ee
where ${\epsilon}_{\pm}$, ${\epsilon}_{0}$ are infinitesimal
parameters
corresponding to the three $SL(2,R)$ generators and
\be
\delta B = 0.
\ee
In this case we have $\delta (\det g^{(\s)}) = 0$ and so eq. (3.25)
implies
\be
\delta \Phi = 0.
\ee
We also conclude from eq. (2.17) that $\delta b = 0$ and so
$\delta \lambda = 0$.

Since $\delta \Phi = 0$, the variation of the metric matrix $g$
is independent of the frame and eqs.
(3.28)--(3.30) can be summarized as follows,
\be
\delta g = \left[ g \left( \begin{array}{cc}
                     0 & 1\\
                     -1& 0  \end{array} \right) ~ , ~
\left( \begin{array}{cc}
{\epsilon}_{0}  & {\epsilon}_{-}\\
{\epsilon}_{+}  &-{\epsilon}_{0}  \end{array}  \right) \right]
\left( \begin{array}{cc}
0  & 1\\
-1 & 0  \end{array}  \right).
\ee
They coincide with the variation (3.15) in the fundamental
representation of $SL(2,R)$. Therefore, in the Einstein
frame the first $SL(2,R)$ subalgebra of the usual $O(2,2)$
string symmetry can be identified with the zero modes of the
first $SL(2,R)$ current subalgebra of the string Geroch group.

(ii). For the other $SL(2, R)$ subalgebra the situation is more
complicated, requiring the use of non--local transformations.
Using the basis elements (3.21), we find that the corresponding
variation (3.27) with infinitesimal parameters
${\epsilon}_{\pm}$, ${\epsilon}_{0}$ yields
\be
\delta g_{AB}^{(\s)} = 2({\epsilon}_{0} + {\epsilon}_{+} B)
g_{AB}^{(\s)},
\ee
\be
\delta B = -{\epsilon}_{-} + 2 {\epsilon}_{0} B + {\epsilon}_{+}
(B^{2} - \det g^{(\s)}).
\ee
We can easily calculate the variation of $\det g^{(\s)}$ and
then use eq. (3.25)
to find the corresponding variation of the dilaton field;
it reads
\be
\delta \Phi = {\epsilon}_{0} + {\epsilon}_{+} B.
\ee
Going to the Einstein frame we find
\be
\delta g_{AB} = 0
\ee
and so the second $SL(2,R)$ subalgebra of the $O(2,2)$ string
symmetry acts only on the $\lambda$--sector, contrary to case
(i), where the situation was the other way around.

We note that $\delta \Phi$ does not depend on ${\epsilon}_{-}$, while
$\delta B$ depends trivially on it (constant $B$--shift). All the
non--trivial dependence of the variations is on ${\epsilon}_{0}$ and
${\epsilon}_{+}$, which we now examine separately. We set first
${\epsilon}_{+} = 0 = {\epsilon}_{-}$ and find that the
axion--dilaton system transforms (up to an overall constant) as
\be
\delta b = -2 {\epsilon}_{0} b, ~~~~ \delta \Phi = {\epsilon}_{0},
\ee
which for the matrix $\lambda$ implies the result
\be
\delta \lambda = \left[ \lambda \left( \begin{array}{cc}
0  & 1\\
-1 & 0  \end{array} \right) ~ , ~ \left( \begin{array}{cc}
{\epsilon}_{0}  &       0\\
0               &-{\epsilon}_{0} \end{array}   \right) \right]
\left( \begin{array}{cc}
0  & 1\\
-1 & 0  \end{array}  \right).
\ee
Comparison with eq. (3.15) shows that in this case the variation of
$\lambda$ can be identified with ${\delta}_{0}^{(0)} \lambda$,
namely the zero mode of the diagonal $U(1)$ subalgebra of the
(other) $SL(2,R)$ current algebra. For ${\epsilon}_{0} = 0 =
{\epsilon}_{-}$, however, the transformation of $\lambda$ is
non--local; we have to go beyond the zero modes to describe it
in a similar way.

It follows from eqs. (3.35) and (3.36) that
\be
\delta({\partial}_{\xi} \Phi) = {\epsilon}_{+} e^{4 \Phi}
\sqrt{\det g}~ {\partial}_{\xi} b, ~~~
\delta({\partial}_{\eta} \Phi) = -{\epsilon}_{+} e^{4 \Phi}
\sqrt{\det g} ~ {\partial}_{\eta} b
\ee
and
\be
\delta(e^{2 \Phi} {\partial}_{\xi} b) = -2 {\epsilon}_{+}
{\partial}_{\xi} (e^{2 \Phi} \sqrt{\det g}), ~~
\delta(e^{2 \Phi} {\partial}_{\eta} b) = 2 {\epsilon}_{+}
{\partial}_{\eta} (e^{2 \Phi} \sqrt{\det g}).
\ee
We claim that this non--local variation of the matrix $\lambda$
can be written in the form (3.16),
\be
\delta \lambda = - \left[ \lambda \epsilon ~ , ~
\left( \begin{array}{cc}
0              & 0\\
{\epsilon}_{+} & 0  \end{array} \right) {\psi}_{\lambda} \epsilon
+ ({{\psi}_{\lambda}}^{t} - 2 {\psi}_{\lambda}) \epsilon
\left( \begin{array}{cc}
0              & 0\\
{\epsilon}_{+} & 0  \end{array}  \right) \right] \epsilon,
\ee
where ${\psi}_{\lambda}$ is the twist potential matrix of
$\lambda$,\footnote{We scale $\lambda$ to
$\sqrt{\det g} ~ \lambda$ in order to apply the same formalism
as was used for the metric sector of the theory. This modification,
however, does not affect eq. (3.42) because $\delta \sqrt{\det g}
= 0$ and so the determinant factor cancels from both sides.}
\be
{\partial}_{\xi} {\psi}_{\lambda} =
\sqrt{\det g} ~ \epsilon~ {(\sqrt{\det g} ~ \lambda)}^{-1}
{\partial}_{\xi} (\sqrt{\det g} ~ \lambda)
\ee
\be
{\partial}_{\eta} {\psi}_{\lambda} =
- \sqrt{\det g} ~ \epsilon~ {(\sqrt{\det g} ~ \lambda)}^{-1}
{\partial}_{\eta} (\sqrt{\det g} ~ \lambda)
\ee
and $\epsilon$ is the antisymmetric matrix (3.3)
as before (not to be confused with the
infinitesimal parameter ${\epsilon}_{+}$ of the variation).
In other words, this transformation is identified with the
$- {\delta}_{-}^{(1)}$ mode of the second $SL(2,R)$ current
algebra of the string Geroch group.
The $SL(2,R)$ algebra of case (ii)
will then be completely described in terms of
the generators ${\delta}_{+}^{(-1)}$, ${\delta}_{0}^{(0)}$ and
${\delta}_{-}^{(1)}$. We recall that
generators with negative modes correspond to trivial transformations,
which explains why the
${\epsilon}_{-}$ part of the variation can be naturally
identified with ${\delta}_{+}^{(-1)} \lambda$ to close the
$SL(2, R)$ algebra.

The proof of eq. (3.42) is straightforward once the right
guess was made. It is sufficient to check the insertion
\be
\delta e^{2 \Phi} = 2 {\epsilon}_{+}
{\psi}_{\lambda}^{11} e^{2 \Phi}, ~~~~
\delta (b e^{2 \Phi}) = 2 {\epsilon}_{+}
{\psi}_{\lambda}^{21} e^{2 \Phi},
\ee
where (11) and (21) denote the corresponding elements of the
$2 \times 2$ matrix ${\psi}_{\lambda}$. On the other hand, the
defining relations (3.43) and (3.44) imply
\be
{\partial}_{\xi} {\psi}_{\lambda}^{11} = \sqrt{\det g} ~
e^{4 \Phi} {\partial}_{\xi} b, ~~~
{\partial}_{\eta} {\psi}_{\lambda}^{11} = - \sqrt{\det g} ~
e^{4 \Phi} {\partial}_{\eta} b
\ee
and
\be
{\partial}_{\xi} {\psi}_{\lambda}^{21} = \sqrt{\det g} ~
(e^{4 \Phi} b {\partial}_{\xi} b - 2 {\partial}_{\xi} \Phi) -
{\partial}_{\xi} \sqrt{\det g},
\ee
\be
{\partial}_{\eta} {\psi}_{\lambda}^{21} = -\sqrt{\det g} ~
(e^{4 \Phi} b {\partial}_{\eta} b - 2 {\partial}_{\eta} \Phi)
+ {\partial}_{\eta} \sqrt{\det g}.
\ee
Then, using eqs. (3.46)--(3.48) it can be verified that
the variation (3.45) coincides with (3.40) and (3.41) after
differentiating ${\psi}_{\lambda}$ with respect to $\eta$ and
$\xi$.

Summarizing the results so far, we have found
that the usual $O(2,2)$
string symmetries are part of a much larger group of transformations
generated by the $O(2,2)$ current algebra. The embedding of
$O(2,2)$, however, is not the obvious one and requires going beyond
the zero modes in the axion formalism of the problem.
In particular, the first $SL(2,R)$ subalgebra
corresponds to ${\delta}_{+}^{(0)}$, ${\delta}_{0}^{(0)}$ and
${\delta}_{-}^{(0)}$, while the second one
corresponds to ${\delta}_{+}^{(-1)}$
${\delta}_{0}^{(0)}$ and ${\delta}_{-}^{(1)}$ in the mode expansion
of the two $SL(2,R)$ current algebras
into which the string Geroch group
is decomposed. In both cases, the variation of the conformal
factor $f$ follows immediately from eq. (3.24) after transforming
it to the Einstein frame.

It is natural to query at this point the
meaning of the global symmetry generated by the zero modes of the
second $SL(2,R)$ current algebra.
Since it only acts on the
$\lambda$--sector of the theory, its infinitesimal form will be
\be
{\delta}^{(0)} \lambda = \left[\lambda \epsilon ~ , ~
\left( \begin{array}{cc}
{\epsilon}_{0} & {\epsilon}_{-}\\
{\epsilon}_{+} & - {\epsilon}_{0} \end{array} \right)
\right] \epsilon,
\ee
in terms of an arbitrary element of the $SL(2,R)$ algebra.
The transformation (3.49) can be equivalently stated as
\be
\delta \Phi = {\epsilon}_{0} + {\epsilon}_{-} b
\ee
and
\be
\delta b = {\epsilon}_{+} - 2 {\epsilon}_{0} b - {\epsilon}_{-}
(b^{2} - e^{-4 \Phi}).
\ee
Then, in terms of the $S_{\pm}$ variables (2.8) that provide an
alternative $\s$--model description of the $\lambda$--sector,
this transformation reads
\be
\delta S_{\pm} = {\epsilon}_{+} - 2 {\epsilon}_{0} S_{\pm}
- {\epsilon}_{-} {S_{\pm}}^{2}.
\ee
It is straightforward to verify that this is the infinitesimal
form of the global transformation
\be
S_{\pm} \rightarrow {A S_{\pm} + B \over C S_{\pm} + D}, ~~~~
AD - BC = 1
\ee
that leaves the $S$--part of the effective action (2.9) invariant.
The relevant infinitesimal expansion is $A = 1 - {\epsilon}_{0}$,
$B = {\epsilon}_{+}$, $C = {\epsilon}_{-}$ and
$D = 1 + {\epsilon}_{0}$, which is an alternative description of
the Lie algebra elements.
The $SL(2,R)$ symmetry (3.53) has been encountered before in 4--dim
string theory, using different methods,
and it is known as $S$--duality [10]. We have just
demonstrated that it is also part of the string Geroch group.

There are many more $SL(2,R)$ subalgebras that can be extracted
from the present formalism. More precisely, the generators
${\delta}_{+}^{(n)}$, ${\delta}_{0}^{(0)}$, ${\delta}_{-}^{(-n)}$
satisfy the commutation relations
\be
[{\delta}_{+}^{(n)} ~ , ~ {\delta}_{-}^{(-n)}] =
{\delta}_{0}^{(0)}, ~~~~
[{\delta}_{0}^{(0)} ~ , ~ {\delta}_{\pm}^{(\pm n)}] =
\pm 2{\delta}_{\pm}^{(\pm n)}
\ee
for any integer $n$, but they are realized non--locally for
$n \neq 0$. Transformations of this type can be applied equally well
to the $g$ or $\lambda$ sectors of string theory, but their
physical interpretation for arbitrary
$n$ lies beyond the scope of the present work.
More generally, the full string Geroch group can be used
to continuously generate infinitely many solutions
descending from any given string background. It is not clear if
its action on the space of solutions is transitive. If this were
the case, any solution could be obtained
(at least formally) from the trivial one in an appropriately
chosen coordinate system. In section 5 we will be
mostly concerned with the construction of multi--soliton solutions,
while the more general problem will be left open to future work.

\section{\bf A Discrete Symmetry of 4--dim Strings}
\setcounter{equation}{0}
\noindent
The 1--loop string background equations in $M_{4}$ with two
commuting space--like Killing fields and orthogonal transitivity
exhibit an additional $Z_{2}$ symmetry.
This manifests itself by interchanging
the field content of the two $SL(2,R)/U(1)$ Ernst $\s$--models that
describe the $g$ and $\lambda$ sectors of the theory. We
consider the transformation
\be
\left( \begin{array}{ccc}
g_{22} & & g_{23}\\
       & &       \\
g_{23} & & g_{33} \end{array} \right) ~ \leftrightarrow ~
\sqrt{\det g} ~ e^{2 \Phi}
\left( \begin{array}{ccc}
1   &  & b\\
    &  &  \\
b   &  & b^{2} + e^{-4 \Phi} \end{array}  \right),
\ee
which is legitimate in the Einstein frame because it leaves the
reduced string equations invariant, without changing the conformal
factor. Indeed, since
\ba
\mbox{Tr} \left( {(g^{-1} \partial g)}^{2} + {({\lambda}^{-1}
\partial
\lambda)}^{2} \right) & = & \mbox{Tr} \left( {(g^{-1} \partial
g)}^{2} +
{\left( {(\sqrt{\det g} ~ \lambda)}^{-1} \partial (\sqrt{\det g}
{}~ \lambda) \right)}^{2} \right) \nonumber\\
& - &
2 {\left( \partial (\log \sqrt{\det g}) \right)}^{2},
\ea
it can be readily verified that the discrete transformation (4.1)
has no effect on $f$ (cf. eqs. (2.32) and (2.33) or their
generalization). Also, $\sqrt{\det g}$ remains invariant.

In $M_{4}$ with signature $-+++$, both $\s$--models are Euclidean.
Therefore, it makes perfect sense to perform the field
interchange and construct the ``mirror image" of any solution
there is available. The space--time interpretation of the
solutions, however, will be quite different from their
``mirror images". We also note that
this transformation is not
an element of the usual $O(2,2)$ group, in general, since
otherwise the dilaton field would
have to stay invariant as well (cf. eq.
(3.24) in the Einstein frame). If we were considering strings
in $M_{4}$ with one space--like and one time--like Killing symmetry,
this field interchange would not be physically correct; in this
case the $\s$--model $g$ would be Lorentzian, while the
$\s$--model
$\lambda$ would stay Euclidean. Similarly, if we were considering
string propagation in 4--dim Euclidean space with two Killing
symmetries, the $\s$--model $g$ would be Euclidean and the
$\lambda$ Lorentzian. The only other case that
the field interchange is consistent with the target space
character of the $\s$--models is string propagation in
4--dim spaces with signature $(2,2)$; in this case both $\s$--models
are Lorentzian, provided that there are one space--like and one
time--like Killing symmetries in the theory.

An interesting example of 4--dim strings with two
space--like Killing symmetries
in $M_{4}$ has been constructed in the
literature using conformal field theories techniques [6, 7].
In particular,
the Lorentzian $(SL(2,R) \times SU(2)) / (U(1) \times U(1))$
coset model provides
a 4--dim solution with non--trivial dilaton and axion fields and
with $\delta c = 0$ to first order in ${\alpha}^{\prime}$,
which describes a closed, inhomogeneous, expanding and recollapsing
universe. The metric [7]
\be
{ds}_{(\s)}^{2} = -{(dX^{0})}^{2} + {(dX^{1})}^{2} +
g_{22}^{(\s)} {(dX^{2})}^{2}
+ g_{33}^{(\s)} {(dX^{3})}^{2} ,
\ee
where
\be
g_{22}^{(\s)} = {2 (1- \sin \theta) {(\sin X^{0}  \sin X^{1})}^{2}
\over (1- \cos 2X^{0}  \cos 2X^{1}) + \sin \theta (\cos 2X^{0} -
\cos 2X^{1})},
\ee
\be
g_{33}^{(\s)} = {2 (1+ \sin \theta) {(\cos X^{0}  \cos X^{1})}^{2}
\over (1- \cos 2X^{0}  \cos 2X^{1}) + \sin \theta (\cos 2X^{0} -
\cos 2X^{1})}
\ee
describes the classical geometry of this model in the
$\s$--model frame of string theory.
The coordinates $X^{0}$, $X^{1}$ take values in the close
interval $[0, \pi /2]$ and $\theta$ is a free parameter that
enters in defining the gauged WZW model.
Also, the dilaton field is
\be
\Phi = -{1 \over 2} ~ \log \left( (1 - \cos 2X^{0}  \cos 2X^{1}) +
\sin \theta (\cos 2X^{0} - \cos 2X^{1}) \right),
\ee
while the only non--zero component of the antisymmetric tensor
field is $B_{23} \equiv B$,
\be
B = - {1 \over 2} ~ {(\cos 2X^{0} - \cos 2X^{1}) + \sin \theta
(1- \cos 2X^{0}  \cos 2X^{1}) \over (1- \cos 2X^{0}  \cos 2X^{1}) +
\sin \theta (\cos 2X^{0} - \cos 2X^{1})}.
\ee
The fields depend only on $X^{0}$, $X^{1}$ and the metric
$g_{AB}^{(\s)}$ is diagonal.

As an application,
we are going to use the discrete transformation (4.1) to generate a
new solution, which is the ``mirror image" of this cosmological
string model. We first find the axion field
\be
b = \cos \theta (\cos 2X^{0} - \cos 2X^{1})
\ee
and
\be
\sqrt{\det g} = {1 \over 2}  \cos \theta ~ \sin 2X^{0}  \sin 2X^{1}
\ee
in the Einstein frame.\footnote{$\sqrt{\det g}$
satisfies the 2--dim wave equation as it
should, and the conjugate solution is $\beta = - {1 \over 2} \cos
\theta ~ \cos 2X^{0} \cos 2X^{1}$.}
The solution that is obtained by performing the field interchange
(4.1)
in the Einstein frame has
zero axion field because the original metric (4.3) is diagonal.
Then, the resulting metric--dilaton system can be rotated back to
the $\s$--model frame, where it assumes the following form: the
dilaton
is still logarithmic,
\be
\tilde{\Phi} = {1 \over 2} \left( \log (\tan X^{0}) +
\log (\tan X^{1}) \right) ~ + ~ \mbox{const.},
\ee
but the metric $g_{AB}^{(\s)}$ develops off--diagonal elements
that are proportional to the original axion field (4.8). We have
explicitly
\be
d {\tilde{s}}_{(\s)}^{2} = f_{1}^{(\s)} \left(-{(dX^{0})}^{0} +
{(dX^{1})}^{2} \right) +
f_{2}^{(\s)} \gamma_{AB}^{(\s)} dX^{A} dX^{B},
\ee
where
\be
f_{1}^{(\s)} = \tan X^{0} \tan X^{1}
\left((1 - \cos 2X^{0} \cos 2X^{1}) +
\sin \theta (\cos 2X^{0} - \cos 2X^{1})\right),
\ee
\be
f_{2}^{(\s)} =
{2 \cos \theta {(\sin X^{0} \sin X^{1})}^{2} \over
(1 - \cos 2X^{0} \cos 2X^{1}) + \sin \theta
(\cos 2X^{0} - \cos 2X^{1})}
\ee
and
\be
\gamma_{22}^{(\s)} = 1, ~~~~~ \gamma_{23}^{(\s)} =
\cos \theta (\cos 2X^{0} - \cos 2X^{1}),
\ee
\ba
\gamma_{33}^{(\s)} & = &
{(\cos 2X^{0} - \cos 2X^{1})}^{2} +
{(1 - \cos 2X^{0} \cos 2X^{1})}^{2} \nonumber\\
& + & 2 \sin \theta (\cos 2X^{0} -
\cos 2X^{1})(1 - \cos 2X^{0} \cos 2X^{1}),
\ea
up to an overall normalization factor $(1 - \sin \theta)/ \cos
\theta$
coming from the constant term of the dilaton field (4.10).
Therefore, only $g_{22}^{(\s)}$ remains invariant under the
field interchange (4.1) in this case.

The solution we have just obtained has
a singularity structure that is different
from the original model. Clearly, the space--time points where
a curvature singularity occurs should coincide with the points
that the dilaton field blows up. In the cosmological solution
(4.3)--(4.5) the only curvature singularities are at
$X^{0} = 0 = X^{1}$ (initial singularity)
or $X^{0} = \pi / 2 =
X^{1}$ (final singularity). In the ``mirror image"
solution the dilaton
field (4.10) blows up separately at each boundary point of $X^{0}$,
$X^{1}$, namely at $X^{0} = 0$ or $X^{1} = 0$ or
$X^{0} = \pi / 2$ or $X^{1} = \pi / 2$. At these points
$\sqrt{\det g}$ also vanishes, as can be readily verified from
eq. (4.9). If we had chosen to work with the special coordinate
system  (2.28), (2.29) in the Einstein frame, with $\sqrt{\det g}$
playing the role of $X^{0}$, the singularities would have
appeared at
$X^{0} \equiv \alpha = 0$.
It is not clear to us whether the ``mirror image"
of the Lorentzian $(SL(2,R) \times SU(2))/(U(1) \times U(1))$
coset model admits a conformal field theory description as well.

It would also be interesting to study the intertwining
of the string Geroch group with the additional discrete
symmetry that interchanges the field content of the two
Ernst $\s$--models. This might be necessary for finding
the right algorithm to generate (at least formally) all
4--dim string backgrounds with two commuting space--like
Killing symmetries. Much work remains to
be done in this direction.

\section{\bf Multi--soliton Solutions}
\setcounter{equation}{0}
\noindent
In this section we describe a special class of solutions,
called multi--solitons, using the integrability
properties  of the dimensionally reduced string
background equations. String multi--solitons are
characterized by two positive integers
$(n,m)$ and descend as a family from any given string background.
Their construction requires the use of the spectral parameter
that appears in the zero curvature formulation of the
reduced string equations. Hence, they correspond to special
elements of the full string Geroch group lying beyond the
range of the usual $O(2,2)$ or $S$--duality transformations.
Soliton calculations in string theory can
be carried out in exact analogy with reduced vacuum Einstein
gravity by
applying the Belinski--Sakharov method to either of the two
$SL(2,R)/U(1)$ Ernst $\s$--models. The pair $(n,m)$ will
therefore refer to the soliton numbers corresponding to the $g$ and
$\lambda$ sectors of the theory. Once this
construction is done in the Einstein
frame, the conformal factor $f$ can be calculated by a simple
integration.

We briefly review the construction of solitons in the Ernst
$\s$--model, since the method is not widely known to string
theorists.\footnote{Soliton methods have been applied before in
de Sitter space--time to construct multi--string solutions [18].}
It is convenient for this purpose to adopt an alternative
formulation of the linearization system (3.5), (3.6) without
making use of the twist potential matrix, but introducing
derivatives with respect to the spectral parameter $l$.
Following Belinski and Sakharov [14], we introduce the differential
operators
\be
D_{1} = {\partial}_{\xi} -2({\partial}_{\xi} \alpha)
{l \over l- \alpha} {\partial}_{l} ~ , ~~~~
D_{2} = {\partial}_{\eta} + 2({\partial}_{\eta} \alpha)
{l \over l + \alpha} {\partial}_{l} ~ ,
\ee
where $\alpha \equiv \sqrt{\det g}$ as usual. Their commutativity,
$[D_{1} ~ , ~ D_{2}] = 0$, is a consequence of the wave equation
${\partial}_{\eta} {\partial}_{\xi} (\sqrt{\det g}) = 0$. Then, the
following system of equations
\be
D_{1} \Psi = {A \over l- \alpha} \Psi, ~~~~ D_{2} \Psi =
{B \over l+ \alpha} \Psi,
\ee
where $\Psi (\eta, \xi; l)$ is a complex matrix function ($l$ can be
complex) and
\be
A = - \alpha {\partial}_{\xi} g {g}^{-1}, ~~~~ B = \alpha
{\partial}_{\eta} g {g}^{-1},
\ee
linearize the Ernst $\s$--model as eq. (3.1) provides their
compatibility condition. The matrix $\Psi$ is a generalization of
$g$ with a spectral parameter so that
\be
g(\eta, \xi) = \Psi(\eta, \xi; l=0).
\ee

Suppose now that a solution $g_{0}(\eta, \xi)$ is already known and
${\Psi}_{0} (\eta, \xi; l)$ is the corresponding solution of the
linear system (5.2). Any other solution descending from it has
\be
\Psi (l) = \chi(l) {\Psi}_{0}(l),
\ee
where $\chi(\eta, \xi; l)$ is a matrix satisfying the equations
\be
D_{1} \chi = {1 \over l- \alpha} (A \chi - \chi A_{0}), ~~~~
D_{2} \chi = {1 \over l+ \alpha} (B \chi - \chi B_{0}).
\ee
$A_{0}$ and $B_{0}$ are the currents (5.3) of the ``seed" matrix
$g_{0}$. We also have
\be
g(\eta, \xi) = \chi(l=0) g_{0}
\ee
and therefore finding $\chi$ will
produce new solutions of the model.
Note that $\chi$ has to satisfy the following additional conditions:

\noindent
(a)  $\bar{\chi} (\bar{l}) = \chi(l)$, so that $\chi$ is real on the
real axis of the complex $l$--plane.

\noindent
(b) $g = \chi ({\alpha}^{2} / l) g_{0} {\chi}^{t} (l)$, with
$\chi (l = \infty) = 1$, so that $g$ is symmetric.

\noindent
(c) $\det \chi(l=0) = 1$, so that $\sqrt{\det g} = \sqrt{\det g_{0}}
=
\alpha$. If this condition is not satisfied, we simply have to
rescale $g$ by $\alpha / \sqrt{\det g}$ so that the determinant
will remain invariant in the coordinate system where the
calculations are performed.

Soliton solutions correspond to the special ansatz for
$\chi$,
\be
\chi(\eta, \xi; l) = 1 + \sum_{k=1}^{n}
\left( {R_{k}(\eta, \xi) \over l- p_{k}(\eta, \xi)} +
{{\bar{R}}_{k} (\eta, \xi) \over l- {\bar{p}}_{k} (\eta, \xi)}
\right),
\ee
up to an overall normalization, which will be determined from the
condition (c); $R_{k}$, ${\bar{R}}_{k}$ are the residue matrices
and $p_{k}$, ${\bar{p}}_{k}$ the locations of the poles in the
complex $l$--plane. They can all be determined explicitly by
substituting the expression (5.8) in eq. (5.6) and comparing the
pole structure of the left and right--hand sides [14]. Setting
$l=p_{k}$, we first determine the location of the poles,
\be
p_{k} = c_{k} - \beta - \sqrt{{(c_{k} - \beta)}^{2} -
{\alpha}^{2}} ~ ,
\ee
where $c_{k}$ are arbitrary complex constants and $\beta$ is the
conjugate variable of $\alpha$ (cf. eq. (2.27)). Then,
\be
q_{k} = c_{k} - \beta + \sqrt{{(c_{k} - \beta)}^{2} -
{\alpha}^{2}} = {{\alpha}^{2} \over p_{k}}
\ee
determine the location of the poles of the inverse matrix
${\chi}^{-1}(l)$. Furthermore, $R_{k}(\eta, \xi)$ turn out to be
degenerate $2 \times 2$ matrices whose elements depend on the
2--vectors
\be
V_{A}^{(k)} (\eta, \xi) = c_{B}^{(k)} {\Psi}_{0}^{-1}
{(\eta, \xi; l= p_{k})}_{BA},
\ee
where $c_{B}^{(k)}$ are arbitrary constant vectors.

For example, for $n=1$, explicit calculation yields
\be
R_{1}{(\eta, \xi)}_{AB} = {1 \over \Delta} \left({V_{D}^{(1)}
{\bar{V}}_{E}^{(1)} {(g_{0})}_{DE} \over q_{1} - {\bar{p}}_{1}}
{\bar{V}}_{C}^{(1)} {(g_{0})}_{CA} -
{{\bar{V}}_{D}^{(1)} {\bar{V}}_{E}^{(1)} {(g_{0})}_{DE} \over
{\bar{q}}_{1} - {\bar{p}}_{1}} V_{C}^{(1)} {(g_{0})}_{CA}
\right) V_{B}^{(1)},
\ee
where
\be
\Delta = {{\mid V_{A}^{(1)} V_{B}^{(1)} {(g_{0})}_{AB} \mid}^{2}
\over {\mid q_{1} - p_{1} \mid}^{2}} -
{{\mid V_{A}^{(1)} {\bar{V}}_{B}^{(1)} {(g_{0})}_{AB} \mid}^{2}
\over {\mid {\bar{q}}_{1} - p_{1} \mid}^{2}}.
\ee
The new solution has $\det g =
{\alpha}^{6} / {\mid p_{1} \mid}^{4}$, which should be rescaled
according to the condition (c) above. The final result
for the (physical) metric $g$ reads
\be
g = {{\mid p_{1} \mid}^{2} \over {\alpha}^{2}} \left(1-
{R_{1} \over p_{1}} - {{\bar{R}}_{1} \over {\bar{p}}_{1}}
\right) g_{0}
\ee
and describes the double--soliton solution on a given
background $g_{0}$. It is not uniquely fixed, however,
as it clearly depends on three complex parameters $c_{1}$,
$c_{B}^{(1)}$. Even in this simple case, $\chi(0)$ is highly
non--trivial, depending on
$g_{0} = \Psi_{0}(l = 0)$ and $\Psi_{0}(l = p_{1})$.

The only non--trivial step in the
calculation is finding
${\Psi}_{0} (\eta, \xi ;l)$ by solving the system (5.2) for
$g = g_{0}$. $\Psi_{0}$ can be determined recursively by
introducing a power series  expansion
\be
\Psi_{0}(\eta, \xi ; l) = \Psi_{0}^{(0)}(\eta, \xi) +
\sum_{k = 1}^{N} l^{k} \Psi_{0}^{(k)}(\eta, \xi),
\ee
where $\Psi_{0}^{(0)} = g_{0}$. This series typically
terminates at a finite integer $N$, depending on $g_{0}$.
Then, the system of differential equations (5.2) yields the
following recursive relations,
\be
{\partial}_{\xi} {\Psi}_{0}^{(k)} =
\left(\alpha {\partial}_{\xi}
+ 2(k+1) ({\partial}_{\xi} \alpha)
+ A_{0} \right) {\Psi}_{0}^{(k+1)},
\ee
\be
{\partial}_{\eta} {\Psi}_{0}^{(k)} = -
\left( \alpha {\partial}_{\eta}  +
2(k+1) ({\partial}_{\eta} \alpha) - B_{0} \right) {\Psi}_{0}^{(k+1)},
\ee
for all $k \geq 0$. It is clear that for finite $N$, the
leading coefficient
${\Psi}_{0}^{(N)}$ is a constant $2 \times 2$ matrix.
The result from the
iteration of the recursive relations (5.16), (5.17) cannot
be brought into a closed form for general $g_{0}$. Therefore,
${\Psi}_{0}(\eta, \xi ; l)$ has to be determined separately in
each case.

Once this is done, any multi--soliton solution
can be constructed explicitly by straightforward generalization
of the $n=1$ case. For $n > 1$, the new solutions of the
Ernst $\s$--model will depend on $3 n$ free complex parameters.
Their form, however, becomes considerably more complicated for
arbitrary $n$. This solution generating technique can be applied
to either of the two $\s$--models
sectors of string theory
in the Einstein frame. As a result, an $(n,m)$ family of
multi--soliton solutions can be constructed explicitly on any
given string background, provided that $\Psi_{0} (\eta, \xi ; l)$
has been determined in each sector. The conformal factor of the
soliton solutions, as has already been pointed out, can be found
by simple integration, in analogy with reduced gravity.
The dressing formula (5.7) has the advantage of providing us
with explicit formulas for the soliton solutions.
The difficult part is finding which particular element of the
Geroch group corresponds to $\chi(0)$, if one wishes to do so.

We note that the ansatz (5.8) describes double, quadruple, etc.,
soliton solutions for $n = 1, 2$, etc. The reason we are not
considering single, triple, etc., soliton solutions is that their
space--time interpretation is problematic. For the single
soliton solution, for example, both $c_{1}$ and $p_{1}$ in eq.
(5.9) have to be real. In this case we are in the region of
space--time with ${(c_{1} - \beta)}^{2} \geq {\alpha}^{2}$;
$p_{1}$ becomes complex
in the complementary region ${(c_{1} - \beta)}^{2} < {\alpha}^{2}$
where the continuation of the
single--soliton solution is a double--soliton with
the special property ${\mid p_{1} \mid}^{2} = {\alpha}^{2}$.
It is well known in this case that the solution (5.14) will remain
unperturbed, i.e. $g = g_{0}$, because the poles of $\chi (l)$
are situated on the circle ${\mid l \mid}^{2} = {\alpha}^{2}$ in
the complex $l$--plane [14]. The point now is that although the
single
soliton solution can be defined continuously everywhere, its first
derivatives will be discontinuous at the points
${(c_{1} - \beta)}^{2} = {\alpha}^{2}$, which is considered to be
problematic.

Many non--trivial solutions of reduced
gravity have been constructed explicitly
by applying the soliton technique to various backgrounds [14].
These results can also be generalized to 4--dim string theory with
non--trivial dilaton and axion fields. We may consider
as an application the $(1, 0)$ soliton solution on the
cosmological background (4.3)--(4.7). Going to the Einstein frame,
where our methods are applicable, we find the following
simple result
for the $g$--sector of the model:
\be
{\Psi}_{0} (X^{0}, X^{1} ;l) = g_{0}(X^{0}, X^{1}) - {l \over
\cos \theta} \left( \begin{array}{ccc}
1- \sin \theta & & 0\\
  &  &  \\
0 & & 1+ \sin \theta \end{array} \right);
\ee
$\alpha$ and $\beta$ have been determined in section 4 and therefore
all the quantities (5.9)--(5.13) can be explicitly found.
Substituting $p_{1}$, $R_{1}$ and their complex conjugates in eq.
(5.14), while keeping the $\lambda$--sector untouched, we obtain
the metric of the $(1,0)$ soliton in the same coordinate system.
One may similarly work out the $(0,1)$ soliton solution on the
``mirror image" background (4.10)--(4.15). The final
expressions will be omitted for simplicity. We point out,
nevertheless, that the matrix $R_{1}$
(and hence $g$ of the $(1,0)$ soliton) has non--zero
off--diagonal elements.
This also implies that the $(0,1)$ soliton solution on the
``mirror image" background carries a non--trivial axion field.

More general solitons can be constructed on this cosmological
background by finding ${\Psi}_{0} (\eta, \xi ; l)$ of
the $\lambda$--sector as well, but the final result is
much more complicated.
A detailed analysis of the corresponding $(n,m)$ soliton
solutions and their space--time interpretation will be presented
elsewhere. It is also conceivable that the Nappi--Witten
universe can be itself
interpreted as a soliton solution around a much
simpler string background, but this possibility will not be
explored here.

It is very difficult to solve the system of equations (5.6) outside
the soliton sector. However, some other
classes of solutions have
been constructed in the literature using different methods.
It is worth mentioning that
meron--like solutions also exist for the Ernst $\s$--model [19],
in analogy with ordinary non--linear
$\s$--models [20]. This last class is
parametrized by two arbitrary functions $f_{1}(\eta)$ and
$f_{2}(\xi)$, which essentially
describe the general solution of the  2--dim
wave equation ${\partial}_{\eta} {\partial}_{\xi} (\sqrt{\det g})
= 0$. The meron--like solutions of the Ernst $\s$--model are
of the form
\be
g = {1 \over \sqrt{f_{1}f_{2}}}
\left(C_{1} log{f_{1} \over f_{2}} + C_{2}\right)
\left( \begin{array}{ccc}
C_{3}(f_{1}f_{2}+1) &  & f_{1}f_{2} - 1\\
  &  &  \\
f_{1}f_{2} -1 &
& (f_{1}f_{2}+1)/C_{3} \end{array} \right),
\ee
where $C_{1}$, $C_{2}$ and $C_{3}$ are arbitrary constants.

It would be interesting to find under what circumstances these
special classes of solutions admit a conformal field theory
interpretation in string theory.

\section{\bf Conclusions and Discussion}
\setcounter{equation}{0}
\noindent
In this paper we have investigated the target space integrability
of 4--dim string theory with dilaton and axion fields in the
presence of two commuting Killing symmetries. The 1--loop string
background equations simplify considerably in the Einstein frame,
reducing to two $SL(2,R)/U(1)$ Ernst $\s$--models, plus
a linear system of
first--order differential equations for the conformal
factor. The non--local symmetries of the two $\s$--models
combine to the string Geroch group, which acts on the space of
solutions
generalizing the usual $O(2,2)$ and $S$--duality transformations.
This infinite dimensional symmetry can be used (at least
formally) as a solution generating technique for new non--trivial
string backgrounds. If $\delta c \neq 0$ to first order in
${\alpha}^{\prime}$, this infinite dimensional structure will
not survive. It will be interesting to see what happens in this
case.

It might be possible to obtain all the solutions from the trivial one
by exponentiating the infinitesimal action of the
underlying $O(2,2)$ current algebra. This expectation does not
contradict the uniqueness theorem of section 2. One might
naively think
that the Geroch group cannot generate non--trivial
solutions from flat space because its action keeps $\sqrt{\det g}$
invariant. This would certainly be true if we had chosen a
coordinate system with $\det g = 1$ in Minkowski space.
We note, however, that the action of
the string Geroch group does not commute, in general, with
coordinate transformations. Therefore,
starting from Minkowski space with
an appropriately chosen coordinate system, so that $\det g$ is not
constant, non--trivial solutions can emerge.
A typical example that illustrates this point is the use of polar
coordinates in 4--dim Minkowski space. Indeed, if we define new
coordinates $(X^{1}, X^{2}) \rightarrow (r, \varphi)$,
\be
X^{1} = r ~ \sin \varphi, ~~~~
X^{2} = r ~ \cos \varphi,
\ee
the metric will become
\be
{ds}^{2} = - {(dX^{0})}^{2} + {dr}^{2} + r^{2} {d \varphi}^{2}
+ {(dX^{3})}^{2},
\ee
which has non--trivial determinant.
The real difficulty lies in making
use of the Geroch group for the explicit construction of new
solutions from old ones. The identification
of group elements is not straightforward even for the
multi--solitons which are much simpler to describe in the
Belinski-Sakharov formalism. More work is certainly required in this
direction.

Another related issue is the problem of boundary conditions.
Infinite symmetries in integrable systems do not necessarily
respect the boundary conditions. Their use for generating new
solutions normally introduces various free parameters,
as in the multi--soliton case. Some of them can be uniquely
determined by imposing the right boundary conditions, but some
arbitrariness may remain. This is physically interesting because
if non--trivial solutions are going to emerge from flat space
(or other simple backgrounds) their characteristic parameters
will have to emerge as well. It would be very desirable to prove
uniqueness theorems for various 4--dim string backgrounds, in
analogy with reduced gravity.

The present results can be generalized to include
electromagnetism (or any number of
additional $U(1)$ fields) in the string background equations.
It is well known that the reduced electrovacuum Einstein
equations exhibit a hidden $SL(3,R)$ symmetry [12, 15] generalizing
the Geroch group of the pure vacuum case. In the presence of
$n$ $U(1)$ fields the corresponding group is
$SL(2+n, R)$.
The appropriate generalization to string theory should
be described by the $O(2, 2+n)$ current algebra, but the
details have to be worked out. Generalizing the results to the
supersymmetric case is also an interesting problem, which could
clarify the role of the infinitely many string symmetries.
In fact, our results might have a more natural interpretation
in terms of hidden symmetries in supergravity theories [21].

We have demonstrated that the non--local
symmetries of $\s$--models can be used in string theory
to enlarge the known $O(2,2)$ and $S$--duality
transformations to an infinite dimensional group. It would be
interesting to formulate the action of the full string Geroch
group in the $\s$--model frame using the antisymmetric tensor
field instead of the axion. Non--local
symmetries can become local and vice versa
in these variables, because $b$ and $B$ are non--locally related
to each other.
Whether this symmetry survives at higher orders in
${\alpha}^{\prime}$ is an important open question. It might
be that only $N = 4$ superconformal theories with $\delta c = 0$
to all orders in ${\alpha}^{\prime}$ [6]
allow for this possibility.
Also, the presence of local symmetries in the axion--dilaton
formalism of string theory
has not been addressed in this paper.
The construction of local conservation laws
in target space might help in clarifying further the
integrability aspects of 4--dim string backgrounds.
Furthermore, the
world--sheet interpretation of the infinitely many
target space symmetries is lacking for
the moment and it should be considered separately.

Finally, the group of duality transformations
$O(2,2;Z)$ might have a natural
generalization in the context of the string Geroch group. If
this is indeed the case, our present understanding of unbroken
string symmetries will improve considerably.

\vskip 2cm
\centerline{\bf Acknowledgements}
\noindent
I am grateful to E. Kiritsis for arising my interest in the
subject and for many useful discussions in the course of this work.
I have also benefited
from conversations with E. Abdalla, L. Alvarez--Gaum\'{e},
C. Kounnas and E. Verlinde.

\newpage
\centerline{\bf REFERENCES}
\begin{enumerate}
\item C. Callan, D. Friedan, E. Martinec and M. Perry, Nucl. Phys.
\underline{B262} (1985) 593.

\item R. Myers, Phys. Lett. \underline{B199} (1987) 371;
I. Antoniadis, C. Bachas, J. Ellis and D. Nanopoulos,
Nucl. Phys. \underline{B328} (1989) 117;
R. Brandenberger and C. Vafa, Nucl Phys.
\underline{B316} (1988) 391;
B. Greene, A. Shapere, C. Vafa and S.--T. Yau, Nucl. Phys.
\underline{B337} (1990) 1;
M. Mueller,
Nucl. Phys. \underline{B337} (1990) 37.

\item E. Witten, Phys. Rev. \underline{D44} (1991) 314;
S. Elitzur, A. Forge and E. Rabinovici, Nucl. Phys.
\underline{B359} (1991) 581; G. Mandal, A. Sengupta and
S. Wadia, Mod. Phys. Lett. \underline{A6} (1991) 1685;
G. Horowitz and A. Steif, Phys. Lett. \underline{B258}
(1991) 91; G. Horowitz and A. Strominger, Nucl. Phys.
\underline{B360} (1991) 197;
J. Horne and G. Horowitz, Nucl. Phys. \underline{B368}
(1992) 444; Phys. Rev. \underline{D46} (1992) 1340; J. Horne,
G. Horowitz and A. Steif, Phys. Rev. Lett. \underline{68}
(1992) 568; D. Garfinkle, G. Horowitz and A. Strominger,
Phys. Rev. \underline{D43} (1992) 3140;
M. McGuigan, C. Nappi and S. Yost, Nucl. Phys.
\underline{B375} (1992) 421;
N. Ishibashi, M. Li
and A. Steif, Phys. Rev. Lett. \underline{67} (1992) 3336.

\item I. Bars and K. Sfetsos, Phys. Rev. \underline{D46} (1992) 4495,
\underline{D46} (1992) 4510; Mod. Phys. Lett. \underline{A7}
(1992) 1091; Phys. Lett. \underline{B277} (1992) 269,
\underline{B301} (1993) 183; E. Fradkin and V. Linetsky,
Phys. Lett. \underline{B277} (1992) 73;
P. Ginsparg and F. Quevedo, Nucl. Phys.
\underline{B385} (1992) 527;
P. Horava, Phys. Lett. \underline{B278} (1992) 101;
A. Sen, Phys. Lett. \underline{B271} (1991) 295,
\underline{B274} (1992) 34; Phys. Rev. Lett.
\underline{69} (1992) 1006; S. Hassan and A. Sen, Nucl. Phys.
\underline{B375} (1992) 103;
D. Gershon, {\em ``Exact Solutions of 4--d Black Holes in String
Theory"}, preprint TAUP--1937--91, 1992;
S. Giddings, J. Polchinski and A. Strominger, {\em ``Four
Dimensional Black Holes in String Theory"}, preprint
UCSBTH--93--14, 1993;
E. Kiritsis, C. Kounnas and D. Lust, {\em ``A Large Class of
New Gravitational and Axionic Backgrounds for Four--Dimensional
Superstrings"}, preprint CERN--TH.6975/93, 1993.

\item K. Meissner and G. Veneziano, Phys. Lett.
\underline{B267} (1991) 33; Mod. Phys. Lett.
\underline{A6} (1991) 3397; G. Gasperini and G. Veneziano,
Phys. Lett. \underline{B277} (1992) 256;
G. Gasperini, J. Maharana and G. Veneziano, Phys. Lett.
\underline{B296} (1992) 51; J. Maharana, Phys. Lett.
\underline{B296} (1992) 65;
A. Tseytlin and C. Vafa,
Nucl. Phys. \underline{B372} (1992) 443;
A. Tseytlin, Int. J. Mod. Phys. \underline{A7} (1992) 223;
Phys. Rev. \underline{D47} (1993) 3421; Nucl. Phys. \underline{B390}
(1993) 153.

\item C. Kounnas and D. Lust, Phys. Lett. \underline{B289}
(1992) 56; C. Kounnas, {\em ``Four--Dimensional Gravitational
Backgrounds Based on N = 4, C = 4 Superconformal Systems"},
preprint CERN--TH.6799/93, 1993.
\item C. Nappi and E. Witten, Phys. Lett. \underline{B293}
(1992) 309; A. Giveon and A. Pasquinucci, Phys. Lett.
\underline{B294} (1992) 162.

\item K. Kikkawa and M. Yamasaki, Phys. Lett. \underline{B149}
(1984) 357; K. Narain, Phys. Lett. \underline{B169} (1986) 41;
T. Buscher, Phys. Lett. \underline{B194} (1987) 59,
\underline{B201} (1988) 466;
A. Shapere and
F. Wilzcek, Nucl. Phys. \underline{B320} (1989) 669;
A. Giveon, E. Rabinovici and G. Veneziano,
Nucl. Phys. \underline{B322} (1989) 167;
A. Giveon, N. Malkin and E. Rabinovici, Phys. Lett.
\underline{B220} (1989) 551;
A. Giveon, Mod. Phys. Lett. \underline{A6} (1991) 2843;
E. Kiritsis, Mod. Phys. Lett. \underline{A6} (1991) 2871.

\item M. Rocek and E. Verlinde, Nucl. Phys. \underline{B373}
(1992) 630; A. Giveon and M. Rocek, Nucl. Phys.
\underline{B380} (1992) 128;
J. Maharana and J. Schwarz, Nucl. Phys. \underline{B390}
(1993) 3;
E. Kiritsis, Nucl. Phys. \underline{B405} (1993) 109;
A. Giveon and E. Kiritsis, {\em ``Axial Vector Duality as a
Gauge Symmetry and Topology Change in String Theory"},
preprint CERN--TH.6816/93, 1993;
S. Hassan and A. Sen, Nucl. Phys. \underline{B405}
(1993) 143; M. Henningson and C. Nappi, Phys. Rev.
\underline{D48} (1993) 861;
L. Alvarez--Gaume, {\em ``Aspects of Abelian and Non--Abelian
Duality"}, preprint CERN--TH.7036/93, 1993; E. Alvarez,
L. Alvarez--Gaume, J. Barbon and Y. Lozano, {\em ``Some Global
Aspects of Duality in String Theory}, preprint CERN--TH.6991/93,
1993.

\item A. Font, L. Ibannez, D. Lust and F. Quevedo, Phys. Lett.
\underline{B249} (1990) 35.

\item A. Einstein and N. Rosen, J. Franklin Inst.
\underline{223} (1937) 43; A. Kompaneets, Sov. Phys. JETP
\underline{34} (1958) 659;
V. Belinski and I. Khalatnikov, Sov. Phys. JETP
\underline{30} (1970) 1174.

\item R. Geroch, J. Math. Phys. \underline{13} (1972) 394;
W. Kinnersley, J. Math. Phys. \underline{18} (1977) 1529;
W. Kinnersley and D. Chitre, J. Math. Phys. \underline{18} (1977)
1538, \underline{19} (1978) 1926, \underline{19} (1978) 2037;
C. Hoenselaers, W. Kinnersley and B. Xanthopoulos,
Phys. Rev. Let. \underline{42} (1979) 481; J. Math. Phys.
\underline{20} (1979) 2530.

\item D. Maison, Phys. Rev. Lett. \underline{41} (1978) 521;
J. Math. Phys. \underline{20} (1979) 871;
B. Harrison, Phys. Rev. Lett. \underline{41} (1978) 1197;
G. Neugebauer, J. Phys. \underline{A12} (1979) L67;
C. Cosgrove, J. Math. Phys. \underline{21} (1980) 2417;
P. Mazur, J. Phys. \underline{A15} (1982) 3173; Acta Phys.
Polon. \underline{B14} (1983) 219;
P. Breitenlohner and D. Maison, Ann. Inst. H. Poinc.
\underline{46} (1987) 215.

\item V. Belinski and V. Sakharov, Sov. Phys. JETP
\underline{48} (1978) 985, \underline{50} (1979) 1;
J. Gruszczak, J. Phys. \underline{A14} (1981) 3247.

\item F. Ernst, Phys. Rev. \underline{167} (1968) 1175,
\underline{168} (1968) 1415;
I. Hauser and F. Ernst, Phys. Rev. \underline{D20}
(1979) 362, \underline{D20} (1979) 1783; J. Math. Phys.
\underline{21} (1980) 1126, \underline{21} (1980) 1418.

\item M.--L. Ge and Y.--S. Wu, J. Math. Phys. \underline{24}
(1983) 1187; Y.--S. Wu, Phys. Lett. \underline{A96} (1983) 179.

\item M. Luscher and K. Pohlmeyer, Nucl. Phys. \underline{B137}
(1978) 46; E. Brezin, C. Itzykson, J. Zinn--Justin and
J.--B. Zuber, Phys. Lett. \underline{B82} (1979) 442; L. Dolan,
Phys. Rev. Lett. \underline{47} (1981) 1371;
T. Curtright and C. Zachos, Phys. Rev. \underline{D24}
(1981) 2661; C. Devchand and D. Fairlie, Nucl. Phys.
\underline{B194} (1982) 232;
B.--Y. Hou, M.--L. Ge and Y.--S. Wu, Phys. Rev.
\underline{D24} (1981) 2238; M.--L. Ge and Y.--S. Wu, Phys. Lett.
\underline{B108} (1982) 411; Y.--S. Wu, Nucl. Phys.
\underline{B211} (1983) 160;
V. Sakharov and A. Mikhailov, Sov. Phys. JETP
\underline{47} (1979) 1017;
K. Ueno and Y. Nakamura, Phys. Lett. \underline{B117}
(1982) 208; Y.--S. Wu, Comm. Math. Phys. \underline{90} (1983)
461.

\item F. Combes, H. de Vega, A. Mikhailov and N. Sanchez,
{\em ``Multi--String Solutions by Soliton Methods in De Sitter
Spacetime"}, preprint LPTHE 93--44, 1993.
\item M. Tseitlin, Theor. Math. Phys. \underline{57} (1984) 1110;
V. Chelnokov and M. Tseitlin, Phys. Lett. \underline{A104}
(1984) 329.
\item D. Gross, Nucl. Phys. \underline{B132} (1978) 439.
\item B. Julia, in {\em ``Superspace and Supergravity"}, S. Hawking
and M. Rocek, eds., Cambridge University Press, Cambridge, 1981;
Lect. Appl. Math. \underline{21} (1985) 355; H. Nicolai,
Phys. Lett. \underline{B194} (1987) 402; H. Nicolai and
N. Warner, Comm. Math. Phys. \underline{125} (1989) 369.
\end{enumerate}
\end{document}